\documentclass[journal]{IEEEtran}
\usepackage[utf8]{inputenc}
\usepackage[utf8]{inputenc}
\ifCLASSINFOpdf
  \usepackage[pdftex]{graphicx}
    \usepackage[pdftex]{graphicx}
\usepackage[utf8]{inputenc}
\usepackage{amssymb}
\usepackage{newunicodechar}
\usepackage{float}
\newunicodechar{≥}{\geq}
\newunicodechar{≤}{\leq}
\else
\fi
\usepackage{amsmath}
\usepackage[colorlinks=true, citecolor=blue]{hyperref}

\UseRawInputEncoding
\begin{document}
%
\title{Secure Autonomous Agent Payments: Verifying Authenticity and Intent in a Trustless Environment}
%
%
%

\author{Vivek~Acharya {}
}

%
%

\markboth{Journal of \LaTeX\ Class Files,~Vol.~1, No.~1, Nov~2025}%
{Acharya \MakeLowercase{\textit{et al.}}: Secure Autonomous Agent Payments: Verifying Authenticity and Intent in a Trustless Environment}
%



\maketitle

\begin{abstract}
Artificial intelligence (AI) agents are increasingly capable of initiating financial transactions on behalf of users or other agents. This shift raises a critical unsolved challenge: how to verify the authenticity of an AI agent and the true intent behind its transactions in a fully trustless, decentralized environment. Traditional payment systems assume a human is present to authorize each purchase, but autonomous agent–led payments break this assumption \cite{1_axios2025googleaiagents}. 

In this paper, we present a novel blockchain-based framework that ensures every AI agent-initiated payment is cryptographically authenticated and intent-verified. Our approach leverages decentralized identity (DID) standards (with verifiable credentials) to establish agent identities and delegations, on-chain intent proofs to capture and verify user authorization for transactions, and advanced cryptographic techniques such as zero-knowledge proofs (ZKPs) to preserve privacy while proving policy compliance. We also incorporate secure execution environments (TEE-based attestations) to guarantee the integrity of the agent’s reasoning process. The proposed system design is detailed with a hybrid on-chain/off-chain architecture that creates an immutable audit trail from initial user instruction to final payment. We qualitatively evaluate the security of this approach, demonstrating how it thwarts impersonation, unauthorized transactions, and misalignment of intent. The discussion highlights the value proposition of secure agent-to-agent payments – enabling autonomous commerce with provable trust and compliance. This work contributes a comprehensive solution for AI agent payments in decentralized systems, laying the groundwork for secure, auditable, and intent-aware autonomous economic agents. We conclude by outlining future research directions to further refine authorization mechanisms, multi-agent coordination, and regulatory compliance in agentic payment networks.\end{abstract}


%
\IEEEpeerreviewmaketitle

\section{Introduction}
%
%
%
%
\IEEEPARstart{AI}-driven agents are poised to transform digital commerce and machine-to-machine transactions. These agents can already “shop, book, negotiate, and pay” on behalf of human users or organizations \cite{2_teleswap2025ap2}. For example, a personal AI assistant might autonomously order supplies when inventory is low, or a trading bot might execute trades based on real-time market conditions. However, empowering software agents to initiate payments autonomously presents a fundamental trust problem. In traditional online payments, a human clicking “Buy” on a trusted interface implicitly provides authentication and intent. With autonomous agents, this direct human presence is absent, breaking the usual trust assumptions. How can a payee or a financial network be sure that an AI agent’s payment request is legitimate – that it truly represents an authorized intent of the user and not a rogue or compromised action?

The core challenge is verifying the authenticity and intent behind AI agent–initiated transactions in a trustless environment. Authenticity means the transaction genuinely originates from a legitimate, authorized agent acting on behalf of a particular user or entity – not an impersonator or unauthorized code. Intent means the transaction’s purpose and details align with what the user (the agent’s principal) actually intended or permitted, ensuring the agent isn’t making unauthorized or unintended purchases. In a fully decentralized setting with no central authority to vouch for the agent, these assurances must be established through technological means. The lack of human oversight in agent-led payments raises risks of fraud, mistakes, and misalignment between user instructions and agent actions, necessitating new mechanisms to establish who an agent is and why it is making a payment.

Various industry and research efforts have begun tackling pieces of this problem. For example, Google’s Agent-to-Agent (A2A) initiative introduced an Agent Payments Protocol (AP2) that uses cryptographically signed user mandates to authorize agent payments \cite{2_teleswap2025ap2}. Self-sovereign identity standards like decentralized identifiers (DIDs) and verifiable credentials (VCs) are being applied to give AI agents persistent, verifiable identities and delegation of authority\cite{4_synergetics2025identity}, \cite{5_cheqd2025verifiableai}. Prior works emphasize that AI assistants will need digital IDs and credentials to establish trust in their actions \cite{7_docklabs2025digitalid}. A recent systematization of knowledge \cite{3_romandini2025sok} noted the need for robust accountability mechanisms for autonomous agents on blockchain, given the irreversible nature of on-chain transactions. It highlighted the absence of standards for logging an agent’s decision process and called for linking agent actions to their reasons, which directly relates to capturing user intent behind transactions. In \cite{6_south2025authdelegation}, an OAuth-inspired delegation framework for authorized AI agents was proposed, highlighting the importance of structured permission tokens for agent actions. Secure agent frameworks like Shade on NEAR use Trusted Execution Environments (TEEs) to run agent code in secure enclaves and verify the code’s integrity on-chain, preventing unauthorized code from acting as a valid agent \cite{9_near2025shadeintro} \cite{10_messari2025shade}. Other approaches propose conditional payment escrows for multi-agent collaborations (e.g., Coral Protocol’s session vaults \cite{8_georgio2025coralprotocol}) to ensure funds only release when tasks are completed. However, no unified framework yet combines decentralized identity, intent verification, cryptographic proofs, and secure execution in a fully trustless, on-chain system for autonomous agent payments.

In this paper, we aim to fill this gap by proposing an integrated solution called Trustless Intent Verification for Autonomous Agents (TIVA). TIVA is a blockchain-based framework that ties agent identity to transaction authorization and enforces that each payment initiated by an AI agent is both authentic and aligned with a user-approved intent, all without relying on any centralized trust. The key contributions of this work include:
\begin{itemize}
\item Decentralized Identity for Agents: We formulate a unified protocol linking AI agent identities (DIDs and verifiable credentials) to transaction authorization, ensuring that every agent-initiated payment can be traced to a legitimate identity and delegation. This provides a scalable, interoperable way to establish authenticity.

\item On-Chain Intent Verification: We introduce on-chain intent verification for autonomous payments using verifiable user mandates and smart contract enforcement, guaranteeing that agent-initiated payments reflect genuine user intent and fall within approved parameters.

\item Privacy-Preserving Compliance: We integrate zero-knowledge proofs to enable privacy-preserving policy checks, allowing agents to prove they adhered to user constraints (spending limits, approved targets, etc.) without revealing sensitive details – a novel application of ZK in AI agent governance.

\item Secure Execution Attestations: We incorporate (optionally) trusted execution environment attestations to secure the agent’s internal processes and keys, adding a layer of defense against agent compromise and reinforcing trustlessness even at the code execution level.

\item Comprehensive Architecture /\& Analysis: We provide a detailed system architecture and a qualitative security analysis of the TIVA framework, bridging gaps noted in prior research such as the need for immutable audit logs and accountability in AI agent operations. Our evaluation demonstrates mitigation of major risks (impersonation, unauthorized spending, fraud) and illustrates the value of provable trust in autonomous agent payments.
\end{itemize}
\section{Methodology}
Research and development at the intersection of AI To address the challenge of verifying agent-initiated transactions in a trustless way, we design a blockchain-based authorization and intent verification protocol for AI agents. Our solution combines several components – decentralized identity, on-chain intent enforcement, cryptographic proofs, and secure execution – into the TIVA framework. Figure 1 provides a high-level overview of the architecture, which we describe below (left side represents the off-chain AI agent, right side the on-chain components).

\textbf {Decentralized Identity and Credentials:} Each AI agent is provisioned with a decentralized identity (DID) and associated cryptographic credentials that define its authorized permissions. When a user or organization creates an agent, they issue a delegation credential (formatted as a W3C Verifiable Credential) attesting to the agent’s identity and what it is allowed to do. For example, the credential might state: “Agent X is authorized to spend up to 5 ETH per day from account Y for the purpose of cloud services.” This credential is signed by the user’s private key, making it tamper-evident and verifiable by anyone using the issuer’s public DID document. The credential can include attributes like spending limits, permitted payees or categories, and expiration times. These credentials are generally stored off-chain for efficiency, but a cryptographic digest or ID can be anchored on-chain (in a DID registry or credential registry) to allow global discovery and to check for revocation. Validating a presented credential involves checking the issuer’s signature (using the issuer’s public key from the DID registry) and confirming the credential has not been revoked. This DID+VC mechanism ensures any party can trustlessly verify an agent’s identity and delegated authority – when an agent presents a signed credential, we know which user or organization it represents and what scope of actions it’s authorized for. If the user revokes or updates the delegation, that status is recorded (e.g., via an on-chain revocation registry) to prevent old credentials from being abused.
\textbf {On-Chain Intent Verification:} Possessing a valid credential gives an agent general authority, but each specific transaction an agent attempts must be checked against the user’s intended purpose for that action. To achieve this, TIVA introduces the concept of an Intent Proof, which serves as a transaction-specific authorization record binding the payment to a prior user instruction. We support two complementary modes for generating intent proofs:
\begin{itemize}
\item Pre-Signed Mandates: The user pre-authorizes a specific task or transaction by signing a structured intent mandate ahead of time. For example, a user could sign a mandate saying “Agent X may buy up to 3 units of item Z at ≤ \$100 each from vendor Y before DD/MM/YYYY.” This mandate (essentially a verifiable credential or signed message) contains the conditions and scope of approval, and is stored off-chain or with the agent. When the agent later wants to execute the purchase, it attaches this signed mandate as proof of user authorization. A smart contract (e.g., a dedicated Agent Wallet Contract) on the blockchain verifies the mandate’s signature (ensuring it was indeed signed by the user) and checks that the transaction details do not exceed the mandate’s constraints (item, quantity ≤3, price ≤\$100, valid vendor, not expired, etc.). Only if the mandate is valid and the transaction falls within its approved parameters will the contract proceed with the payment; otherwise, the payment is rejected as out of scope. This approach is analogous to Google’s AP2 off-chain mandates but implemented entirely on-chain for trustless enforcement. All validation logic lives in the smart contract, eliminating any need for a centralized intermediary to approve the agent’s action.

\item Dynamic On-Chain Policy: For ongoing delegations or flexible tasks, a user may not want to pre-sign every possible intent. Instead, the user can deploy a policy smart contract encoding rules that the agent must follow. For instance, a policy contract could specify: “Agent X can spend up to 10 USDC per day on cloud services; anything beyond requires my manual approval.” When the agent attempts a payment, it calls a function on this policy/escrow contract. The contract autonomously checks the current state (e.g., how much Agent X has spent today) and either authorizes the transfer or blocks it according to the encoded rules. The agent’s identity (DID) is used to ensure only the legitimate agent can invoke this contract on the user’s behalf. Essentially, the policy contract acts as an on-chain gatekeeper or escrow, holding the user’s funds and releasing them only when conditions reflecting the user’s intent are met. This pattern is similar to the Coral Protocol’s session vaults which escrow a task’s budget until completion. By moving intent verification logic on-chain, we gain transparency and automatic enforcement via blockchain consensus.
\end{itemize}

In our architecture, the Agent Wallet Contract combines these roles: it serves as a smart contract wallet that holds the agent’s spending funds (or controls access to the user’s funds) and enforces the above mandate/policy checks on every transaction. The Agent Wallet Contract is programmed such that only the agent’s DID (public key) is authorized to initiate payments from it, and each call must include a valid intent proof (either a user-signed mandate or evidence that the on-chain policy conditions pass). The contract verifies the agent’s signature and the intent proof before executing the transfer. This design sandboxes the agent’s financial actions – the agent cannot arbitrarily spend funds without providing the cryptographic proof that the user intended that specific payment. If any check fails, the contract aborts the transaction. Every authorized payment then produces an immutable on-chain record (event log) linking the transaction to the intent proof for future audit. By using a contract-based wallet with built-in verification logic, we achieve defense in depth: even if an agent is compromised, stolen credentials or malicious code alone cannot transfer funds unless the proper user-signed intent is also present.

\textbf {Zero-Knowledge Proofs for Privacy:} A significant challenge is balancing transparency with privacy. Publishing detailed intent mandates or credentials on-chain could reveal sensitive user information or business logic. To mitigate this, TIVA employs zero-knowledge proofs in the intent verification process to prove compliance with policies without revealing private details. We use ZKPs in two main ways:
\begin{itemize}
\item \textbf{Selective Disclosure of Credentials:} The agent can prove it possesses a valid credential with certain attributes meeting the requirements, without exposing the entire credential. For example, suppose the agent has a credential stating it can spend up to \$500. Instead of revealing this credential or the exact limit, the agent provides a ZKP attesting that “I have a credential issued by User U that authorizes at least \$500 of spending.” The proof convinces the verifier (smart contract or payee) that such a credential exists and is signed by the user, while hiding the actual limit and any other permissions. This selective disclosure ensures the agent’s authorization is validated, but unnecessary details (like other privileges or exact amounts) remain confidential.
\item \textbf{Intent Compliance Proofs:} Similarly, for each transaction, the agent can prove that the transaction complies with the user’s approved intent parameters without revealing those parameters. For instance, if the user’s mandate says “purchase item Z if price ≤ \$100”, the agent can generate a ZK proof that “I have a signed mandate from the user and the price in this transaction does not exceed the mandate’s price limit”. The smart contract verifies this proof (using an on-chain zk-SNARK verifier or an off-chain verification followed by an on-chain validity check) against the user’s public key. The proof shows that (a) the mandate was indeed signed by the user, and (b) the purchase amount is within the allowed limit, without revealing the actual price or limit value. If an agent tries to cheat (e.g., buy something beyond the limit), it cannot produce a valid proof, and the contract will reject the transaction. This approach ensures that even on a public ledger, the details of user intent (which might be sensitive) aren’t exposed, yet the rules are rigorously enforced. Using efficient zk-SNARKs, these proofs can be verified with minimal gas cost, and support for such verification is increasingly available on modern blockchain platforms.
\end{itemize}
By leveraging ZKPs, privacy and compliance go hand-in-hand: agents demonstrate they are following user mandates or policies truthfully, but only the necessary facts are revealed to observers. This prevents leakage of confidential business information or personal data, while still holding the agent accountable to the user’s conditions.

\textbf{Secure Execution Environments:} In addition to on-chain measures, TIVA can optionally leverage Trusted Execution Environments to secure the agent’s off-chain computation. A TEE (such as Intel SGX or ARM TrustZone) allows the agent’s code to run in an isolated, hardware-protected enclave. We utilize TEEs to ensure the agent’s internal decision logic and cryptographic keys cannot be tampered with by an outside attacker or even the host of the agent process. When an agent is running inside a TEE, it can produce a remote attestation – a cryptographic quote from the hardware proving that the agent’s code is genuine and untampered. Our system can verify this attestation on-chain (e.g., via a known attestation contract or oracle) before an agent’s actions are accepted. In practice, this means the Agent Wallet Contract would require the agent to provide an attestation quote (or have a set of whitelisted enclave public keys) along with transactions. Only if the quote is valid – meaning the agent’s code hash matches the approved code and is running in real hardware – will the contract honor the request. This prevents a malicious or modified AI instance from masquerading as the legitimate agent. For example, Shade agents on NEAR employ multiple independent TEEs that must all sign off on a transaction; if one enclave is compromised, the others (with correct code) will refuse, blocking the action. Our framework can accommodate such designs (e.g., threshold signatures from enclaves) to eliminate single points of failure. While using TEEs involves trusting the hardware manufacturer, it adds a robust layer of security at the agent’s source of decisions. Even if TEEs are not available, at minimum the agent should protect its private keys in a secure module (like an HSM or software vault) to reduce the risk of key compromise. By combining secure enclaves with our on-chain verification, we achieve a defense-in-depth model: cryptographic policies guard the outputs of the agent, and TEEs guard the internal process, making it extremely difficult for an attacker to either impersonate the agent or cause it to deviate from the user’s intent.

In summary, the TIVA framework marries off-chain AI autonomy with on-chain trust guarantees. The AI agent can operate flexibly in the off-chain world – interacting with services, gathering information, making decisions – but whenever it seeks to spend funds or complete a payment, the blockchain infrastructure steps in to verify identity, validate intent, and enforce constraints through code, not trust. This yields an end-to-end solution where every autonomous transaction is backed by verifiable evidence of who the agent is and why the transaction is taking place.

\begin{figure} [htbp]
\centering
  \includegraphics[width=\columnwidth]{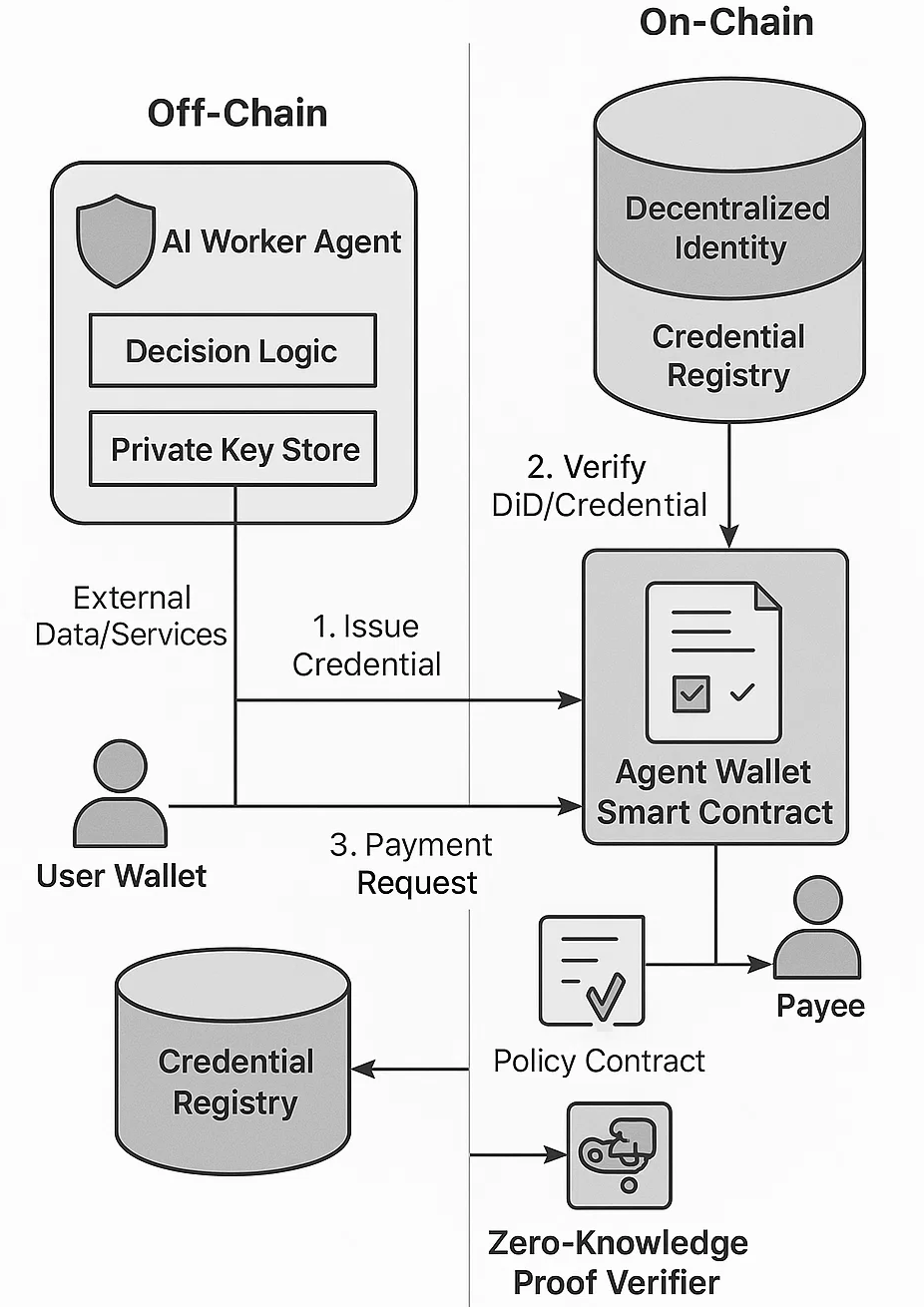}
  \caption{Conceptual architecture of the Trustless Intent Verification for Autonomous Agents (TIVA). 
  The off-chain domain hosts the AI worker agent (optional TEE) for decision logic and key storage; 
  the on-chain domain verifies DID/credentials and enforces payments via the agent wallet smart contract. 
  The user wallet issues verifiable credentials (1), the wallet verifies DID/credential (2), and the agent submits a payment request (3). 
  Optional policy and zero-knowledge proof modules support rule evaluation and privacy.}
  \label{fig:tiva_architecture}
\end{figure}

\section{Results and Security Analysis}
We conducted a qualitative security analysis of TIVA to evaluate how well it meets the requirements of authenticity and intent verification in a trustless setting. The analysis considered various threat scenarios and how our design mitigates them. Key security properties of the TIVA framework are summarized below:
\begin{itemize}
\item \textbf{Agent Authenticity and Impersonation Resistance:} Every transaction is signed by the agent’s private key (tied to its DID), and the corresponding public key is recorded on-chain, so no attacker can impersonate the agent without stealing its keys. By anchoring identities in a decentralized DID registry, we eliminate reliance on a central PKI or certificate authority. As long as the agent’s private key remains secure (especially when protected by a TEE or hardware module), an adversary cannot forge the agent’s identity or valid transaction signatures. Moreover, since the agent’s delegation credential is also signed by the user, an attacker would need to compromise both the agent’s key and the user’s key to create a malicious agent with valid permissions. This dual-key requirement greatly increases security: even if an agent’s key is stolen, the thief gains no authority without the user’s credential; and if the user’s key is stolen, it cannot be used to operate the agent without the agent’s own key. TIVA thus provides strong cryptographic assurance that the agent initiating a payment is the genuine one authorized by the user, thwarting impersonation attacks.
\item \textbf{Intent Verification and Misuse Prevention:} The core promise of our framework is that an AI agent cannot initiate a payment that the user did not intend or permit. This is enforced through the mandate and policy checks: any payment request must include a user-signed intent or satisfy a pre-defined on-chain policy. If an agent tries to exceed its mandate (e.g., spend more than allowed or pay an unapproved recipient), the smart contracts will detect the discrepancy and reject the transaction. By tying each action to explicit cryptographic proof of intent, TIVA prevents an agent from misbehaving beyond the limits of what’s technically enforceable. While an agent could still act within its allowed scope in undesirable ways (a risk mitigated by careful mandate design), it cannot violate the quantifiable constraints set by the user. This greatly reduces the risk of unauthorized or rogue transactions occurring without the user’s consent.
\item \textbf{Trustless Decentralization:} The system is designed to be fully trustless, meaning no centralized authority or intermediary is required to approve agent transactions. All verification (identity, signatures, proofs) is handled by smart contracts running on the blockchain, and trust is placed in well-vetted cryptographic primitives and distributed consensus. There is no need to trust the agent itself or any third-party service; even the user who created the agent cannot secretly manipulate transactions without the on-chain checks catching it. This decentralization ensures the approach is robust even in open, permissionless environments with potentially adversarial participants.
\item \textbf{Accountability and Auditability:} Every agent action produces an immutable audit trail on the blockchain. Each transaction record is linked to the evidence of user authorization (e.g., includes a reference or hash of the intent mandate). This provides strong non-repudiation – neither the user nor the agent can later deny that a particular transaction was authorized. In the event of a dispute (say an unexpected charge), the merchant or auditor can point to the on-chain proof that the agent had a mandate for that payment, or conversely the user can show if an agent acted without a valid mandate. The transparent logs enable post-hoc analysis and forensic investigation of agent behavior, which is critical for building trust in autonomous systems. Our design essentially brings the agent’s “intent logs” into the open: instead of hidden reasoning, there is at least a public record of the intended reason for each payment.
\item \textbf{Privacy Preservation:} Despite the transparency of verification, the use of zero-knowledge proofs means sensitive details (like exact spending limits, item descriptions, or business policies) need not be revealed on-chain. We achieve policy enforcement without broadcasting the policy itself. For example, outside observers might see that “Agent X paid 50 USD to Service Y with proof of authorized intent” but not know the user’s original spending cap or the specifics of the mandate. This optional privacy layer is important for real-world adoption, as it protects user confidentiality and commercial secrets while still proving compliance. In essence, TIVA offers provable compliance with user intent, without requiring full transparency of that intent.
\end{itemize}
In addition to these points, our analysis considered the attack surface and failure modes. We found that many classic attack vectors are addressed by the framework’s multi-layered approach. For instance, credential forgery is infeasible due to digital signatures (a fake mandate or delegation will fail verification against the issuer’s public key). Key compromise remains a concern (as with any crypto system), but the impact is limited by defense-in-depth: a stolen agent key alone cannot authorize new actions without the user’s signature, and vice versa. If both keys were stolen, the attacker essentially becomes the user, which is beyond the scope of agent-specific protections (it reduces to standard private key security). The use of TEEs further mitigates key theft and unauthorized code execution, as discussed. We also ensure that an agent’s privileges can be instantly revoked by the user (by marking its credential as revoked on-chain), which stops any further transactions from that agent once a compromise is detected. Overall, the TIVA framework significantly raises the security bar for autonomous agent payments: any attacker must defeat multiple independent safeguards (identity verification, intent proof checks, enclave attestation, etc.) to maliciously move funds, making the system resilient against a wide range of threats.
\section{Discussion}
The proposed secure intent-verification framework for AI agent payments brings several important benefits and broader implications for the future of autonomous systems:
\begin{itemize}
\item \textbf{User Empowerment and Control:} TIVA allows users to confidently delegate payment tasks to AI agents while retaining ultimate control at a policy level. Users can specify high-level spending rules or mandates for their agents instead of micromanaging every transaction. This granular consent mechanism means the user’s intent is always explicit and enforceable. If the user’s goals or risk tolerance change, they can update or revoke credentials and mandates at any time, and the changes take effect immediately for all future agent actions. Such a framework empowers users to harness agent automation (saving time and effort) without sacrificing the ability to rein in the agent – the agent becomes an accountable extension of the user, not an uncontrollable risk.
\item \textbf{Fraud Reduction and Merchant Confidence:} Businesses and payees receiving agent-initiated payments benefit from stronger assurances that the payment is legitimate and authorized. When an AI agent pays for an item or service through our system, the merchant not only receives funds but also a verifiable proof of the user’s intent (the attached mandate or on-chain proof). This is stronger evidence of authorization than a traditional credit card transaction, for example. It can drastically reduce fraud and chargeback disputes – a merchant can trust that the agent’s payment was approved by the user, because it’s cryptographically non-repudiable. If a user later claims “I didn’t authorize that purchase,” the merchant can point to the on-chain intent proof signed by that user. Thus, merchants and service providers can process autonomous agent payments with greater confidence, potentially automating fulfillment immediately upon receiving payment + proof (since the order details and authorization are embedded). This could streamline e-commerce by reducing the need for manual order review and fraud screening. Overall, transactional friction is lowered while security is raised, making agent-driven commerce efficient and trustworthy.

\item \textbf{Regulatory and Compliance Alignment:} Our framework provides built-in features that could help satisfy regulatory requirements around digital payments and AI. Every transaction is tied to an identified agent and a user’s explicit authorization, which addresses concerns around anonymous or untraceable AI actions. The immutable audit logs and identity credentials can facilitate compliance with financial regulations such as KYC (Know Your Customer), AML (Anti-Money Laundering), and consumer consent laws. For example, an agent’s DID could be linked to a verified identity, and credentials could encode that an agent is allowed to transact only with whitelisted addresses or under certain limits, in accordance with legal requirements. Smart contracts could automatically enforce spending limits or sanctions lists by design. Moreover, from an AI governance perspective, many ethical frameworks call for transparency and user control over AI systems. TIVA’s design provides both: a transparent record of actions and cryptographic user control over what agents can do. If future regulations mandate that “AI agents must clearly signal the scope of their authority and maintain records of their actions,” our solution is essentially a technical implementation of that principle. Thus, adopting such a system could help developers and organizations future-proof their autonomous agent services against emerging legal and ethical standards.
\item \textbf{Interoperability and Standardization:} We intentionally built TIVA on open standards like DIDs and VCs, and designed it to complement emerging protocols like A2A and AP2. This fosters interoperability in a broader ecosystem of autonomous agents. For instance, if another project issues AI agent identity tokens as NFTs or uses different DID methods, our Agent Wallet could recognize and work with those credentials as long as they are verifiable. Similarly, our on-chain intent verification approach could integrate with or inspire extensions to standards like AP2. By demonstrating how an on-chain intent-proof layer can work, this research can influence standardization efforts, encouraging industry consortia to include similar mechanisms in their specifications. The goal is a future where secure, intent-verifying agent transactions are not a proprietary solution but a common capability across platforms. Our contributions (e.g., the idea of zero-knowledge proof of intent compliance or DID-based agent authentication) can be building blocks that others adopt, leading to a more secure and cohesive agent ecosystem at large.
\end{itemize}
In summary, TIVA advances the state of the art for secure autonomous agents by uniting identity, intent and trustless enforcement in one framework. It lays a foundation for provably trustworthy AI agents that can participate in economic activities. By addressing the key trust issues (Who is the agent? Was this action authorized?), we pave the way for broader adoption of autonomous agents in finance and commerce. Users, businesses, and regulators can have greater confidence in agentic transactions when they are backed by verifiable cryptographic evidence of authenticity and intent. This work also opens avenues for future innovation – from developing more expressive yet safely verifiable intent languages, to refining the human-agent interaction for issuing mandates (e.g., intuitive UIs that help users specify their intent correctly), to extending these concepts to multi-agent collaborations with collective decision-making. We believe that as these techniques mature, they will become integral to the infrastructure of decentralized finance and IoT, ensuring that autonomous agents operate in alignment with human objectives and constraints.
\section{Conclusion}
We have presented TIVA, a secure payment framework for autonomous AI agents that verifies both the agent’s identity and the user’s intent behind each transaction in a trustless environment. This work addressed a critical emerging challenge at the intersection of AI, blockchain, and digital payments: how to ensure an AI agent’s financial actions are authentic and authorized in the absence of direct human oversight. Our solution combines decentralized identity (DIDs with verifiable credentials), on-chain smart contracts for intent enforcement, zero-knowledge proofs for privacy, and optional secure hardware execution, to create an end-to-end protocol for autonomous agent payments with provable trust and alignment. In the TIVA model, every transaction an agent makes can be independently verified to have originated from a legitimate agent and to conform to a user-approved intent, with enforcement by code rather than by trusting the agent or any intermediary. We demonstrated through the system architecture and security analysis that this approach can mitigate major risks – such as agent impersonation, unauthorized spending, and fraud – and provide strong accountability for agent behavior.

The introduction of intent-aware, non-repudiable agent transactions has significant implications. It effectively brings concepts from Google’s AP2 and related authorization frameworks into a fully decentralized context, enabling autonomous commerce with the assurance of on-chain compliance checks. The agent is free to act within its delegated scope, but the moment it tries to step outside, the blockchain will catch it. This shifts the paradigm from trusting an AI agent to verifying it. We believe this is a crucial step for scaling AI-driven automation in financial systems – it allows AI agents to participate in economic activities with a high degree of autonomy without compromising security or user control.

In conclusion, this work establishes a foundation for secure, auditable, and intent-aware autonomous agent ecosystems. Future efforts will focus on implementing a TIVA prototype in a live blockchain environment to assess performance, gas costs, and user experience. This involves developing smart contracts for credential and proof verification and creating an AI agent that interacts with them in practice. Further exploration of advanced intent representations, domain-specific policy languages, and multi-agent coordination under joint intent proofs will extend the framework’s utility. We believe privacy-preserving intent verification will inspire continued research at the intersection of AI and blockchain, advancing a future where autonomous transactions are verifiably authentic, aligned with user intent, and seamlessly integrated into decentralized finance and commerce.
\ifCLASSOPTIONcaptionsoff
  \newpage
\fi



%

%
\bibliographystyle{IEEEtran}
\bibliography{bare_jrnl}

\end{document}